\documentclass[12pt]{article} 

\newenvironment{pf}{\noindent{\bf Proof:}}{\newline\kbn}

\usepackage{amssymb,theorem}
\usepackage{german,a4,isolatin1}
\newtheorem{theo}{Theorem}

\newtheorem{prop}[theo]{Proposition}

\newcommand{\Einsop}{\leavevmode{\rm 1\mkern  -4.4mu l}}
\newcommand{\Seins}{{\mathsf{S}^1}}

\newcommand{\PSL}{{\mbox{\it PSL}}}

\newcommand{\komm}[2]{{\left[ #1 , #2 \right]}} 
\newcommand{\betrag}[1]{{\left| #1 \right|}} 
\newcommand{\norm}[1]{{\left\| #1 \right\|}} 
\newcommand{\klammer}[1]{{\left( #1 \right)}}

\newcommand{\lok}[1]{{\mathcal #1}}

\newcommand{\Name}[1]{{\sc #1}} 
\newcommand{\dopp}[1]{{\mathbb #1}}

\newcommand{\opo}{{1\!+\!1}}
\newcommand{\closure}[1]{{#1}^-}
\newcommand{\menge}[1]{{\left\{ #1 \right\}}}

\newcommand{\kbn}{$\square$}

\begin{document}

\title{Absence of stress energy tensor in CFT${}_2$ models}
\author{S\o{}ren K"oster\\Inst.\ f.\ Theor.\ Physik\\ Universit\"at
  G\"ottingen\\ Tammannstr. 1, 37077 G\"ottingen \\Germany}
\date{} 

\maketitle
\begin{abstract} 
We prove: the local quantum theories defined by conformally covariant
derivatives of the $U(1)$-current algebra in $1+1$ dimensions do not
contain a stress-energy tensor in the sense of the theorem of L\"uscher
and Mack. 
\\[1mm]
\end{abstract}

\section{Introduction}
\label{sec:intro}

Much of the present understanding of quantum field theories was
a\-chie\-ved by methods related to internal and space-time
symmetries. There are reasons to be interested in objects connected
with symmetries which are of a local nature, and in this work we
concern ourselves with densities generating specific space-time 
symmetries. 

Within the classical framework the relation between local objects and
continuous symmetries of a \Name{Lagrange}an field theory is canonical
by \Name{Noether}'s theorem: to each such symmetry we have an explicitly
known conserved current, whose integrals over space, the charges,
generate the corresponding symmetry transformation. In quantum field
theory the situation is less satisfactory. If one quantises a classical
\Name{Lagrange}an field theory, it may happen that some symmetries do
not survive at all because of renormalisation effects. Moreover,
there is no a priori knowledge of densities 
connected with continuous symmetries of a general
quantum field theory, although it is
possible to characterise such fields abstractly, of course. The nature
of conserved currents connected 
to symmetries at the quantum level (and of their charges in
particular) is hard to clarify in general. 

These problems are more accessible for the global conformal
space-time symmetry in $1+1$ dimensions. Here we have an abundance of
models for which explicit constructions of a conserved \Name{Wightman}
quantum field are known, which serves as a density for the conformal
symmetry. When smeared out with suitable test functions,  this field
actually generates the conformal symmetry in the sense of
integrable \Name{Lie} algebra representations. Its interpretation as a
stress-energy tensor is in direct analogy with the classical object. 

Depending on weak assumptions \Name{L\"uscher and Mack}
found that stress-energy tensors of conformally covariant quantum
field theories in $1+1$ dimensions always yield local formulation of the
\Name{Virasoro} algebra \cite{FST89} (theorem 3.1). 

We prove: No such stress-energy tensor exists
in a class of completely 
well-behaved conformal theories in $1+1$ dimensions, the conformally
covariant derivatives of the $U(1)$ current. These
are constructed as fields on
\Name{Minkowski} space and possess  conformally 
covariant extensions on their own \Name{Fock} space, but
they do not transform covariantly with respect to the transformations
implementing global conformal symmetry of the $U(1)$ current.

\Name{Yngvason}
\cite{jY94} studied the conformally covariant derivatives as part of a  
broader class of derivatives of the $U(1)$ current and  established,
among other things, 
that they do not fulfill \Name{Haag}
duality on \Name{Minkowski} space\footnote{Essential duality, which
  is another name for \Name{Haag} duality on the conformal covering of
\Name{Minkowski} space, is a consequence of conformal symmetry.}. 
\Name{Guido, Longo and Wiesbrock} \cite{GLW98} studied locally normal
representations of these models and found representations of the
first derivative, which do not allow an 
implementation of global conformal symmetry. In a closing side-remark
they noted that this contradicts, by
unpublished results of \Name{D'Antoni and Fredenhagen}, presence of
diffeomorphism symmetry 
in these models. A general version of these results is available in
\cite{sK03e}, but at this point we are 
interested in a straightforward argument excluding presence of a
stress-energy tensor for the derivative models.

As a first step we establish the conformally covariant derivatives of
the $U(1)$ current as conformally covariant local quantum theories,
following a method 
introduced by \Name{Buchholz and 
  Schulz-Mirbach} \cite{BS90}. 
 

\section{The $\Phi^{(n)}$-models}
\label{apphi}

The $U(1)$ current in $\opo$ dimensions decomposes into two independent chiral
components, the chiral currents, and we shall discuss one of these
only. The derivatives of the chiral
current $j$ are given as fields on the light-ray by
$\Phi^{(n)}(x) := \partial_x^n j(x)$, where we used
$\partial_x:= d/dx$. These fields are covariant with
scaling dimension $n+1$ when acted upon by the implementation of the
stabiliser group of $\infty$ for the 
$U(1)$-current theory. By looking
at their \Name{Wightman} functions one recognises that the
$\Phi^{(n)}$ possess 
conformally covariant extensions, if restricted to their own
\Name{Fock} space; the corresponding unitary representation of
$\PSL(2,\dopp{R})$ implementing global conformal symmetry leaves
invariant the vacuum and fulfills the spectrum condition. Each of
these extensions transforms covariantly 
with respect to a different representation  of the global conformal
group and lives on a different \Name{Fock} space.
From now on,  we look at the
fields as operators in their cyclic subrepresentation equipped with
their own representation of the global conformal group and we 
use the symbol $\Phi^{(n)}$ in this sense. 

By construction, the derivative fields $\Phi^{(n)}$ obey the following
commutation relation as fields on the light-ray:
$$\komm{\Phi^{(n)}(x)}{\Phi^{(n)}(y)} = \frac{i}{2\pi} (-)^n
\delta^{(2n+1)}(x-y)\Einsop\,\,.$$ 
We want to calculate the corresponding commutation relations for the
modes of the conformally extended fields on the compactified light-ray,
$\Seins$. These fields will be denoted $\widetilde{\Phi}^{(n)}$.  The
test-functions of 
fields on $\Seins$ and their 
images living on the light-ray are connected by a transformation
 $f\mapsto \widehat{f}$ depending on the scaling dimension of
the respective field; its definition is induced by
$\widetilde{\Phi}^{(n)}(f)\equiv \Phi^{(n)}(\widehat{f})$. The inverse
transformation is denoted as $\widehat{f}\mapsto \widetilde{\widehat{f}}=f$.
\begin{prop}\label{prop:Phinmodcomm}
  The modes $\Phi^{(n)}_m:=
\widetilde{\Phi}^{(n)}([z^{n+m}])$ have
the following commutation relations:
\begin{equation}\label{modcomrel}
  \komm{\Phi^{(n)}_m}{\Phi^{(n)}_{m'}} =
 \delta_{m,-m'}
   \Pi^{(n)}(m)\,\, ,
\end{equation}
if we set $\Pi^{(n)}(m):= \prod_{k=0}^{2n} (m-n+k)$.
\end{prop}
Remark: These relations imply that the modes $\Phi^{(n)}_m$,
$|m|\leq n$, are central, which in turn means that all
$L_0$-eigenspaces for eigenvalues $1,\ldots,n$ are null.

\begin{pf}
We use the
shorthand notations $\zeta:=(1+z)$, $d/dz=
\partial_\zeta$ and arrive at:
\begin{eqnarray}
  &&\komm{\widetilde{\Phi}^{(n)}(f)}{\widetilde{\Phi}^{(n)}(g)} \equiv
  \komm{\Phi^{(n)}(\widehat{f})}{\Phi^{(n)}(\widehat{g})}
\nonumber\\ 
&=&(-)^{2n+1} \oint \frac{dz}{2\pi i} f(z)
\zeta^{-2(n+1)}\klammer{\zeta^2\frac{d}{d\zeta}}^{2n+1}
\zeta^{-2n}g(z)\nonumber\\
&=& \oint \frac{dz}{2\pi i} g(z) \klammer{\frac{d}{dz}}^{2n+1} f(z)
  \,\, .
\label{compcomrel}
\end{eqnarray}

The identity of the two integration kernels as distributions may be
proved inductively.  Applying the induction 
assumption we see that we have to prove:
$  \zeta^{-2n} \partial_\zeta \zeta^2\partial_\zeta \zeta^{2n} \partial_\zeta^{2n-1} \zeta^{-2} = \partial_\zeta^{2n+1}$.
One may verify this identity for $n=1$ explicitly
. Then one proves by induction on
$n$: 
\begin{eqnarray*}
&&\zeta^{-2(n+1)} \partial_\zeta \zeta^2\partial_\zeta \zeta^{2(n+1)}
\partial_\zeta^{2n+1} \zeta^{-2}\\
&=&\zeta^{-2}\partial_\zeta^{2n+1}
\klammer{\zeta^2\partial_\zeta^2 -
  2(2n+1)\zeta^2\partial_\zeta \zeta^{-1}+(2n+1)(2n)} =
  \partial_\zeta^{2n+3} \,\, .
\end{eqnarray*}  
\end{pf}

As we can see by looking at their canonical commutation relations,
the derivative fields may be treated as local quantum theories of
bounded operators in terms of \Name{Weyl} operators and their
relations (cf \cite{GLW98}). We take another approach
which was introduced  by \Name{Buchholz and
  Schulz-Mirbach} \cite{BS90} for the nets of the stress-energy tensor
and the $U(1)$ current: By  establishing
linear energy bounds referring to the conformal {Hamilton}ian
$L_0$  the \Name{Haag-Kastler}
axioms follow from \Name{Wightman}'s  set of axioms. 
In particular the
fields are essentially self-adjoint on the 
\Name{Wightman} domain, their bounded functions fulfill locality and
the local algebras generate  a dense subspace from
the vacuum. The local algebras are generated by unitaries $W(f):=
exp(i\closure{\widetilde{\Phi}^{(n)}(f)})$,
$\closure{\widetilde{\Phi}^{(n)}(f)}$ self-adjoint and 
$supp(f)\Subset\Seins$. The $W(f)$ are concrete representations
of the \Name{Weyl} operators. 
\begin{prop}
  The following defines the local algebras of the chiral net generated
  by the $\Phi^{(n)}$ fields:
  \begin{equation}
    \label{eq:deflokalgphin}
    \lok{A}_{\Phi^{(n)}} (I) := \menge{\closure{\widetilde{\Phi}^{(n)}(f)},
  \,\, supp(f) \subset I, f = \widetilde{\overline{\widehat{f}{\,}}}\,\,}'', \,\,
I\Subset \Seins \,\, .
  \end{equation}
\end{prop}

\begin{pf}
The proof follows the lines indicated in \cite{BS90}.
If $\psi_N$ denotes an arbitrary eigenvector of $L_0$ with energy
$N$ and norm $1$, then $\Phi^{(n)}_m\psi_N$, $m>0$, is a multiple of a unit
vector of energy $N-m$, which we call
$\psi_{N-m}$. Making use of a general estimate for positive, linear
functionals $\eta$ \cite{dB90}: $|\eta(Q)|^2\leq \eta(Q^*Q)$, we are led to
the following bound: 
$$
  \|\Phi^{(n)}_m\psi_N\|^4 
\leqslant
\|\Phi^{(n)}_m\psi_N\|^2 \|\Phi^{(n)}_m\psi_{N-m}\|^2+\Pi^{(n)}(m)
\|\Phi^{(n)}_m\psi_N\|^2 \,\, .
$$
 For $m\neq0$ we set $\Pi'{}^{(n)}(m):= \frac{1}{m}\Pi^{(n)}(m)$ and 
we prove inductively using the spectrum condition:
$
\|\Phi^{(n)}_m\Psi_N\|^2 \leq N \Pi'{}^{(n)}(m)$, $ m\geq 1
$.

For the generating modes we have:
$$  \|\Phi^{(n)}_{-m}\Psi_N\|^2= \|\Phi^{(n)}_m\Psi_N\|^2 + \Pi^{(n)}(m) \leq
(N+m) \Pi'{}^{(n)}(m) \,\, .
$$ 
The zeroth mode is central in the theory and is,
therefore, a multiple $q$ of the identity
. So we have: $\|\Phi^{(n)}_{0}\Psi_N\|^2 = q^2$, $q\in\dopp{R}$.   
For general $\Phi^{(n)}(f)$, $f\in C^\infty(\Seins)$, and a vector
$\Psi$ from the \Name{Wightman} domain we have the following estimate:
\begin{eqnarray}
  \norm{\Phi^{(n)}(f)\Psi} &\leqslant& \norm{(L_0+\Einsop)\Psi} \sum_{m\in\dopp{Z}}
  \betrag{f_m}\klammer{\betrag{m}+\Pi'{}^{(n)}(m)+\betrag{q}+1} \,\, .
   \label{hbound}
\end{eqnarray}
This is the linear energy bound from which the \Name{Haag-Kastler} axioms
follow as discussed in \cite{BS90}.
\end{pf}

Now shortly on the nuclearity condition for the conformally covariant
derivatives of the $U(1)$ current. 
Since null vectors reduce the multiplicity of $L_0$
eigenvalues, the trace of $e^{-\beta L_0}$, $\beta> 0$, of the
vacuum representation of the derivative models is
dominated by the $L_0$ character for the $U(1)$ current, which is
given by the combinatorial partition function $p(e^{-\beta})$
. The following discussion applies for the
same reason to all theories defined by a stress-energy tensor and to
the $U(1)$-current algebra. $p(e^{-\beta})$ is directly
connected to \Name{De\-de\-kind}'s $\eta$-\-function:
\begin{displaymath}
  p(e^{-\beta})^{-1} = \prod_{m\geq 1} (1-e^{-\beta m}) =
  e^{-\frac{\beta}{24}}\eta(i\beta/(2\pi))   \,\, .
\end{displaymath}

For the nuclearity condition we have to check the asymptotic behaviour
for $\beta \searrow 0$. This behaviour is determined by the
transformation law of $\eta$ for $\tau \rightarrow 1/\tau$. It reads
\cite{bS74} (III.\S3)
:
$  \sqrt{\beta/2\pi} \,\eta(i\beta/(2\pi)) = \eta(i2\pi/\beta)$.
We have with $\beta_0>-1+\pi^2/6$ and $n=1$:
\begin{equation}
  \label{eq:nuclcond}
   \lim_{\beta \searrow 0} p(e^{-\beta})
   e^{-\klammer{\frac{\beta_0}{\beta}}^n} = 0  \,\, .
\end{equation}
This estimate is a special form of a nuclearity condition and ensures
the split property for 
all models under consideration by arguments as given in 
\cite{FG93} (Lemma 2.12.).


\section{No stress-energy tensor in $\Phi^{(n)}$ models}
\label{sec:prep}

We seek for a stress-energy tensor in the theories defined by
conformally covariant  derivatives of degree $n$ of the $U(1)$ current
in $\opo$ 
dimensions. We assume the stress-energy tensor to deserve its name and
therefore it should be a
local, covariant, conserved, symmetric, traceless quantum field
$\Theta$ of scaling dimension 
$2$, which is relatively local to the $\Phi^{(n)}$ under consideration
and a density for its infinitesimal conformal
transformations. Because all models involved factorise into chiral
components, we shall discuss the situation on the
compactified light-ray, ie the fields  live on
$\Seins$. 

According to the analysis of \Name{L\"uscher and Mack} the commutation
relations of $\widetilde{\Theta}$ have a very 
specific form \cite{LM76,gM88} \cite[theorem 3.1]{FST89}. $\widetilde{\Theta}$ is a 
\Name{Lie} field with an extension proportional to $c$, the {\em
  central charge} of $\widetilde{\Theta}$:
$$\frac{c}{12}
\oint \frac{dz}{2\pi i} f'''(z) g(z) = \komm{\widetilde{\Theta}(f)}{\widetilde{\Theta}(g)} -
\widetilde{\Theta}(f'g-fg')\,\,.$$
$c/2$ is the normalisation constant of the two
point function of $\widetilde{\Theta}$, hence we have $c\in\dopp{R}_+$,
and, by the
\Name{Reeh-Schlieder} theorem, 
 $\widetilde{\Theta}=0$ if and only if  $c=0$. 

\begin{prop}\label{prop:noset}
  The conformally covariant derivatives of the $U(1)$ current in $\opo$
  dimensions do not contain a stress-energy tensor.
\end{prop}

\begin{pf}
Looking at the commutation
relations of the modes of $\Phi^{(n)}$ (equation \ref{modcomrel}), we
learn that the eigenspaces 
of the conformal {Hamilton}ian $L_0$ associated with energy
$1,\ldots,n$ are all null.  If 
$n\geq 2$ this yields for  $L_{-2}=\widetilde{\Theta}([z^{-1}])$: 
$c/2 = \|L_{-2}\Omega\|^2 = 0$, and hence $\widetilde{\Theta}=0$. 

In the case $n=1$ all vectors of energy $2$ are multiples of
$\Phi^{(1)}_{-2}\Omega$. If there is a stress-energy tensor
$\widetilde{\Theta}$, we have: $ \gamma L_{-2} \Omega =
\Phi^{(1)}_{-2}\Omega$, 
$c|\gamma|^2 = 12$. Obviously, 
$\widetilde{\Phi}^{(1)}- \gamma \, \widetilde{\Theta}$ is a
quasi-primary field and 
its two-point function is determined by conformal covariance up to a
constant, $C\geq 0$:
\begin{displaymath}
  \langle \Omega,\, 
(\left.\widetilde{\Phi}^{(1)}\right.^\dagger(z)- \bar{\gamma}
\widetilde{\Theta}^\dagger(z))  
(\widetilde{\Phi}^{(1)}(w)- \gamma \widetilde{\Theta}(w)) \, \Omega \rangle
= 
C (z_> -w)^{-4} \,\, .
\end{displaymath}
In particular, we have:
$
  C = \|(\Phi^{(1)}_{-2}-\gamma
    L_{-2})\Omega\|^2 = 0
$.
By the \Name{Reeh-Schlieder} theorem, the field
$\widetilde{\Phi}^{(1)}- \gamma \widetilde{\Theta}$ is zero. Since
$\gamma^{-1}\widetilde{\Phi}^{(1)}$ is not a 
stress-energy tensor, the claim holds for $n=1$ as well.  
\end{pf}


\section{Discussion}
\label{sec:dis}

We have shown that a quantum field theory of conformally covariant
derivatives of the 
$U(1)$ current in $1+1$ dimensions does not contain a stress-energy
tensor. This adds another detail to their character as archetypes of
conformal theories in $1+1$ dimensions: In spite of being simple and
completely well behaved, they do not exhibit special properties of
other comparatively simple models such as strong additivity or
presence of a stress-energy tensor.

 If there
is a local density associated in some sense with the conformal
symmetry of these models, it has to be of a different nature.

 We mention just one reason  why
such densities are desirable. Looking at a chiral conformal theory
$\lok{B}$ in its vacuum representation and at a covariant subnet
$\lok{A}\subset \lok{B}$, the 
question arises, whether the local relative commutants
$\lok{A}(I)'\cap\lok{B}(I)$ define a subtheory as well. The problem is
to show that the relative commutants increase with $I$, i.e. to prove
isotony for this set. If there is a  
sufficiently well behaved local density for the dilatations in the
globally inner representation $U^\lok{A}$, this can actually be
confirmed.  
This program has been carried out in presence of stress-energy
tensors \cite{sK03c}, but it should be feasible in more general
settings as well: There ought to be sufficiently many local
observables to answer such questions. 

A general quantum version of \Name{Noether}'s theorem exists
\cite{BDL86} on grounds of the split property, which is 
established easily for the   conformally
covariant derivatives (see section \ref{apphi}). Here, symmetries are
implemented 
on local algebras by operators which are localised in a somewhat
enlarged region. These {\em local implementers} are densities
of a different nature than eg stress-energy tensors,
since they define a representation of the respective 
symmetry group with the same spectral properties as the original one,
but it has been established that they provide approximations
for the global implementation of the respective  
symmetry \cite{ADF87}. It is not clear, however, whether these local
implementers will prove 
sufficiently well behaved, if we want to apply them to
the isotony problem
. 

\Name{Carpi} \cite{sC99a} reconstructed the stress-energy tensor
of some models by taking point-like limits of the local implementers
applying methods of \Name{J\"or\ss{} and Fredenhagen} \cite{FJ96} and
hence gave an explicit account of the relation of local
implementers to densities as needed in \cite{sK03c}. 
It appears important to study the local implementers as
densities for the conformal transformations in more detail and
starting with a look at the conformally covariant
derivatives seems to be promising.

\subsection*{Acknowledgements}
I owe much to
\Name{K.-H.Rehren} (G\"ottingen) for helpful discussions
 and a critical reading of the manuscript.
 Financial support from the \Name{Ev.\ Studienwerk
    Villigst} is gratefully acknowledged.



\end{document}